\documentclass[twocolumn,showpacs,amsmath,amssymb]{revtex4}

\usepackage{amsmath}
\usepackage{epsf}
\begin{document}

\title{Nonlinear effects for Bose Einstein condensates\\in optical lattices}
\author{M.~Jona-Lasinio (*), O.~Morsch, M. Cristiani,
E.~Arimondo}
\affiliation{INFM, Dipartimento di Fisica E.Fermi, Universit\`{a}
di Pisa, Via Buonarroti 2, I-56127 Pisa,Italy\\(*) e-mail:
jona@df.unipi.it}
\author{C. Menotti}
\affiliation{Dipartimento di Fisica, Universit\`{a}
di Trento and BEC-INFM, I-38050 Povo, Italy}

\date{\today}

\begin{abstract}
We present our experimental investigations on the subject of dynamical
nonlinearity-induced instabilities and of nonlinear Landau-Zener tunneling
between two energy bands in a Rubidium Bose-Einstein condensate in an
accelerated periodic potential. These two effects may be considered two
different regimes (for small and large acceleration) of the same physical
system and studied with the same experimental protocol. Nonlinearity introduces
an asymmetry in Landau-Zener tunneling; as a result, tunneling from the ground
state to the excited state is enhanced  whereas in the opposite direction it is
suppressed. When the acceleration is lowered, the condensate exhibits an
unstable behaviour due to  nonlinearity. We also carried out a full numerical
simulation of both regimes integrating the full Gross-Pitaevskii equation; for
the Landau-Zener effect we also used a simple two-level model. In both cases we
found good agreement with the experimental results.
\end{abstract}
\pacs{PACS number(s): 03.65.Xp, 03.75.Lm}

\maketitle
\section{Introduction}
Cold atoms and, more recently, Bose-Einstein condensates (BECs) in optical
lattices have attracted increasing interest since their first
realization~\cite{firstbec}. In particular, the formal similarity between the
wavefunction of a BEC inside the periodic potential of an optical lattice and
electrons in a crystal lattice have triggered theoretical and experimental
efforts alike. Many phenomena from condensed matter physics, such as Bloch
oscillations and Landau-Zener tunneling have since been shown to be observable
also in optical lattices~\cite{anderson,morsch01,inguscio04}. In a recent
experiment,  a BEC in an optical lattice even made possible the observation of
a quantum phase transition that had, up to then, only been theoretically
predicted for condensed matter systems~\cite{greiner02}. However, an important
difference between electrons in a crystal lattice and a BEC inside the periodic
potential of an optical lattice is the strength of the self interaction and
hence the magnitude of the nonlinearity of the system. Electrons in a metal are
almost noninteracting whereas atoms inside a BEC interact strongly. A
perturbation approach is appropriate in the former case while in the latter the
full nonlinearity must be taken into account. From this feature new physics is
expected. Most experiments to date have been carried out in the regime of
shallow lattice depth, for which the system is well described by the mean field
Gross-Pitaevskii equation with a periodic potential. Moreover, the nonlinearity
induced by the mean-field of the condensate has been shown, both theoretically
and experimentally, to give rise to
instabilities~\cite{wu01,wu03,machholm04,chiara03,scott04,modugno04,
cristiani04,fallani04,javanainen04} in certain regions of the Brillouin zone. These
instabilities are not present in the corresponding linear system, i.e. the
electron system.

In this paper we review and summarize our experimental and theoretical results
on the subject of nonlinear Landau-Zener tunneling and nonlinearity-induced
instabilities in a Bose-Einstein condensate interacting with an external
periodic potential. These two phenomena represent the most dramatic
manifestations of nonlinearity in two different regimes of the system. In order
to study these phenomena we have used a single experimental procedure. The
underlying idea is to linearly scan the Brillouin zone, by applying a constant
acceleration to the periodic potential, and cross the band edge. Then the
condensate is released and an absorption picture of the condensate is taken
after a time of flight, reflecting the momentum distribution at the time of
release. By varying the nonlinearity of the system with a fixed (large)
acceleration, we can study the nonlinear contributions to the Landau-Zener
effect. We shall denote this regime as the ``Landau-Zener'' regime. On the
contrary, by varying the acceleration from very small to intermediate values
with a fixed nonlinearity, we can study the stability of the condensate and the
effects of nonlinearity on the dynamics. We will refer to this regime as the
``instability'' regime.

This paper is organized as follows. After describing our theoretical approach
in section~\ref{theory}, we explain our experimental techniques in
section~\ref{experiment}. Section~\ref{landauzener} presents a discussion of
our results on the Landau-Zener tunneling, and the experimental and conceptual
difficulties encountered in obtaining them. For the interpretation of  the
nonlinear Landau-Zener tunneling we re-examine and critically compare the
effective  potential concept, introduced into previous investigations, with our
present results. Furthermore we interpret the asymmetry in the nonlinear
Landau-Zener effect on the basis of different chemical potentials calculated
for the ground band and for the first excited band. Section~\ref{instab}
discusses our experimental and theoretical  results on the condensate
instabilities.  Finally our conclusions and perspectives for future
developments and improvements are given in section~\ref{conclusions}.

\begin{figure}[htbp]
\centering\begin{center}\mbox{\epsfxsize 3.4 in \epsfbox{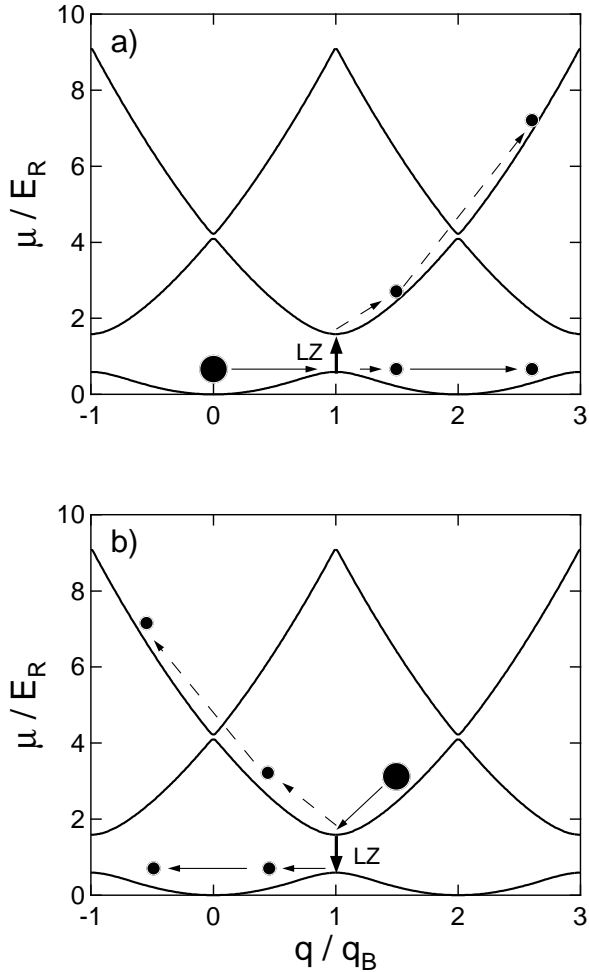}}
\caption{Schematic representation of the band structure for the chemical
potential $\mu$ versus the quasimomentum $q$ in an optical lattice ($s=2$) and
LZ-tunneling (ground to excited band (a) and excited to ground band (b)). When
the BEC is accelerated across the edge of the Brillouin zone (BZ) at
quasimomentum $q_{B}=\hbar k_{L}$, LZ tunneling can occur. Further acceleration
will result in the condensate part in the upper level, to undergo LZ tunneling
to higher bands with a large probability (due to the smaller gaps between
higher bands) being that part essentially unaffected by the lattice. After the
first crossing of the BZ edge, we increase the lattice depth and decrease the
lattice acceleration thus reducing the tunneling rate from the ground band to
the excited band to much lower values at successive BZ edge
crossings.}\label{Band-structure}
\end{center}\end{figure}

\section{Theory}
\label{theory}
The motion of a Bose-Einstein condensate in an accelerated 1D optical lattice
is described by the Gross-Pitaevskii equation
\begin{multline} \label{schrod}
i\hbar \frac{\partial \psi}{\partial t} =
\frac{1}{2M}\left(-i\hbar\frac{\partial}{\partial
x}-Ma_{L}t\right)^2\psi +  \\ + \frac{V_0}{2}\cos(2k_lx)\psi +
\frac{4\pi\hbar^2a_s}{M}\left|\psi\right|^2\psi
\end{multline}
where $M$ is the atomic mass, $k_L=\pi/d$ is the optical lattice wavenumber
with optical lattice step $d$.  $V_0$ is the periodic potential depth,
$E_{R}=\hbar^2 k_L^2/2M$ is the recoil energy. We introduce the dimensionless
parameter $s$
\begin{equation}
     V_{0}=s E_{R}
\end{equation}
denoting the lattice depth in units of the recoil energy. The $s$-wave
scattering length $a_s$  determines the nonlinearity of the system, with the
two-body coupling constant given by $4\pi\hbar^{2}a_{s}/M$.
Equation~\eqref{schrod} is written in the comoving frame of the lattice, so the
inertial force $Ma_L$ appears as a momentum modification. The wavefunction
$\psi$ is normalized to the total number of atoms in the condensate and we
define $n_0$ as the average uniform atomic density.  By defining the
dimensionless quantities $\tilde x=2k_Lx$, $\tilde  t=8E_{R}t/\hbar$,
$\tilde{\psi} =\psi/\sqrt{n_0}$, $ v=s/16$, $\alpha=M a_L/16E_{R}k_L$,
$q_B=1/2$. The nonlinearity is characterized through the
parameter~\cite{choi99}
\begin{equation}
C=\frac{\pi n_0 a_s}{k_L^2}.
\end{equation}
Therefore  eq.~\eqref{schrod} is cast in the following 
form~\cite{wuniu-lz,jona1,jona2}:
\begin{multline} \label{schrod-adim}
i\frac{\partial \psi}{\partial t} = \frac{1}{2}\left(
-i\frac{\partial}{\partial x}-\alpha t \right)^2 \psi +v
\cos(x)\psi + C \left|\psi \right|^2 \psi
\end{multline}
where we have replaced $\tilde x$ with $x$, etc. In the neighborhood of the
Brillouin zone edge, at quasimomentum $q = q_{B}$,  we   approximate the wave
function by a superposition of two plane waves with complex coefficients  (the
two level model illustrated in~\cite{wuniu-lz}), assuming that only the ground
state and the first excited state are populated~\cite{note1}. We then
substitute in eq.~\eqref{schrod-adim}
\begin{equation} \label{twolevels}
\psi (x,t)=a(t)e^{iqx}+b(t)e^{i(q-1)x},
\end{equation}
with $|a(t)|^2+|b(t)|^2=1$. Comparing the coefficients of $e^{iqx}$ and
$e^{i(q-1)x}$, linearizing the kinetic terms and dropping the irrelevant
constant energy $1/8+C[1+(|a|^2+|b|^2)/2]$, eq.~\eqref{schrod-adim}  assumes
the form
\begin{multline}
\label{two-state-eq}
i\frac{\partial}{\partial t}\,\begin{pmatrix} a \\ b \end{pmatrix} =
\left[\alpha t\frac{\sigma_{3}}{2}+v\frac{\sigma_{1}}{2}\right]
\begin{pmatrix} a \\ b \end{pmatrix} + \\
-\frac{C}{2} \begin{pmatrix} (|a|^2-|b|^2) & 0 \\ 0 & -(|a|^2-|b|^2) 
\end{pmatrix}
\begin{pmatrix} a \\ b \end{pmatrix}
\end{multline}
where $\sigma_i\,\,(i=1,2,3)$ are the Pauli matrices. Each solution of
eq.~\eqref{two-state-eq} has an associated conserved energy $\epsilon[\psi]$
\begin{multline}
\epsilon[\psi]=\alpha t <\psi|\frac{\sigma_3}{2}|\psi>+v
<\psi|\frac{\sigma_1}{2}|\psi>+ 
\\ -C \left(<\psi|\frac{\sigma_3}{2}|\psi>\right)^2
\quad \mbox{with } |\psi> \equiv \begin{pmatrix} a \\ b \end{pmatrix}
\end{multline}

It must be stressed that the energy and the hamiltonian eigenvalue do not
coincide, since we are dealing with a nonlinear system. The energy is the
quantity that is conserved along the system trajectories while the hamiltonian
eigenvalue is not. In nonlinear systems the hamiltonian eigenvalue is usually
called the chemical potential (indicated with $\mu$). The connection between
energy $\epsilon$ and chemical potential $\mu$  for  a condensate within an
optical lattice was discussed in~\cite{kraemer03}. Those quantities  are
equivalent in the case of negligible atomic interactions, i.e. $C \sim 0$. The
Bloch bands may describe either the energy or the chemical potential as a
quasimomentum function. In the following we choose to consider the Bloch bands
as the curves representing the {\it chemical potential} $\mu$  as a function of
the quasimomentum (fig.~\ref{Band-structure}).

The adiabatic Bloch bands of eq.~\eqref{two-state-eq} have a swallow tail
structure (i.e. they develop a loop and become multi-valued) at the edge of the
Brillouin zone for $C \ge v$~\cite{wuniu-lz,pethick,mueller,machholm03}, a
regime not explored  in our experiments.

\begin{figure}[htbp]
\centering\begin{center}\mbox{\epsfxsize 3.2 in
\epsfbox{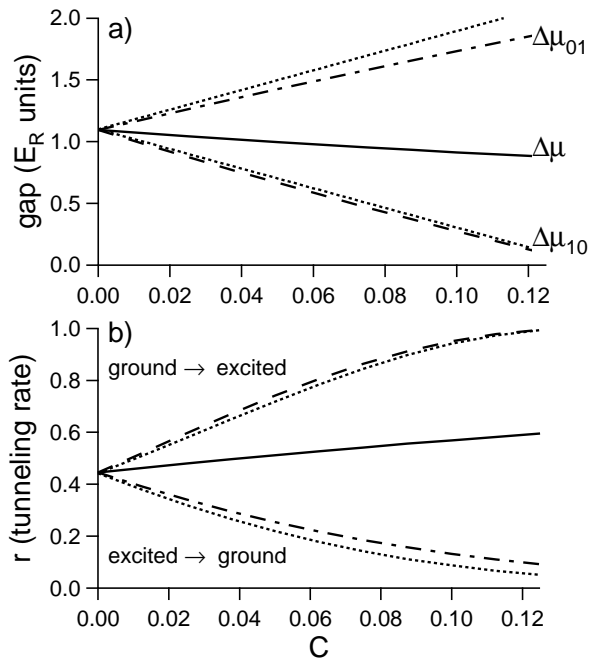}}
\caption{In a), the energy gaps between the ground and the excited Bloch bands
calculated at the BZ edge for different values of the nonlinear parameter $C$
are shown. The continuous line represents the energy gap $\Delta \mu$ while the
dashed and dot-dashed lines shows the energy gaps $\Delta \mu_{10}$ and $\Delta
\mu_{01}$ as explained in the text. The dotted lines are
eq.~\eqref{effectmattia} for both tunneling directions. In b), the Landau-Zener
tunneling rate $r$ calculated from the energy gaps in a) and from
eq.~\eqref{effectmattia}, with the same conventions
for the linestyles as in a), is shown. The calculations were done for a lattice
depth $s=2.2$. }\label{gap}
\end{center}\end{figure}

In fig.~\ref{gap}(a) we plot the band gap at the BZ edge 
between the lower band and the first excited band in the nonlinear case,
calculated in regimes that are relevant for the Landau-Zener experiments.
We defined the gap in the non linear case as the chemical potential
difference between the two bands at the edge of the BZ. However a
definition of this kind is manifold: we define $\Delta \mu$ as the
difference in chemical potential between the first excited and ground
band, both calculated self-consistently assuming the total density
of the condensate in each of them; moreover we define $\Delta \mu_{10}$ as the
chemical potential difference between the first excited and
ground band, both consistent with the total density in the ground
band (i.e. using the ground state
wavefunction in the mean field term) and $\Delta \mu_{01}$ as the chemical
potential difference
between the first excited and ground band, both consistent with
the total density in the first excited band (i.e. using the excited state
wavefunction in the mean field term). In order to describe a
transition from the ground band to the upper one, we believe that
the gap $\Delta \mu_{10}$ is better suited, and conversely the gap $\Delta
\mu_{01}$ should better describe the tunneling in the opposite
direction.

\section{Experimental setup} \label{experiment} Our experimental apparatus for
creating BECs of ${^{87}}\mathrm{Rb}$ atoms was described 
in~\cite{muller00}. The main feature of our apparatus relevant for the present
work is the triaxial time-averaged orbiting potential (TOP) trap with trapping
frequencies $\nu_x:\nu_y:\nu_z$ in the ratio $2:1:\sqrt{2}$. Our trap is,
therefore, almost isotropic.
The optical lattice is created by two laser beams with parallel linear
polarizations and wavelength $\lambda$, as described 
in~\cite{cristiani02}. The two beams are derived from the first diffraction
orders of two acousto-optic modulators that are phase-locked but with
independent frequencies, allowing us to introduce a frequency difference
$\Delta \nu$ between them. The resulting periodic potential has a variable
lattice constant $d$ depending on the intersecting angle between the two laser
beams. The smallest lattice constant, $0.39 \mathrm{\mu m}$, is obtained in
the counterpropagating configuration and can be increased up to $1.2
\mathrm{\mu m}$ when the two lasers intersect at about $38$ degrees. The
depth $V_{0}$ of the periodic potential (depending on the laser intensity and
detuning from the atomic resonance of the rubidium atoms) can be varied from
$0\,E_{R}$ up to approximatively $3\,E_{R}$. It must be noted that $E_{R}$ is
the {\it true} recoil energy and depends on the angle at which the two laser
beams intersect. The beams are detuned to the red side of the rubidium atomic
resonance by $30\,\mathrm{GHz}$. In this way, a periodic potential with lattice
recoil energy $E_{R}/h = 455\,\mathrm{Hz}$ is created. In addition, by
linearly chirping the frequency difference $\Delta \nu$, the lattice is
accelerated with $a_{L}=d\frac{d\Delta\nu}{dt}$. In our experiments, we used 
accelerations ranging from $a_{L}=0.3\,\mathrm{m\,s^{-2}}$ to
$a_{L}=5\,\mathrm{m\,s^{-2}}$.

The experimental protocol for `moving' the condensate across the Brillouin zone
is as follows. After creating BECs with roughly $10^4$ atoms, we adiabatically
relax the magnetic trap frequency to $\nu_x=42\,\mathrm{Hz}$. Thereafter, the
intensity of the lattice beams is ramped up from $0\,E_{R}$ to a value
corresponding to a lattice depth of approximatively $2\,E_{R}$. Once the final
lattice depth is reached, the lattice is accelerated for a time $t$. Finally,
both the magnetic trap and the optical lattice are switched off, and the
condensate is observed by absorption imaging after a time-of-flight of
$21\,\mathrm{ms}$.

\section{Landau-Zener tunneling}
\label{landauzener}
\subsection{Modelling}
Evaluating the transition probability in the adiabatic approximation for the
transition from an initial state to a final one separated by an energy gap,
we find the linear LZ formula for the tunneling probability $r$
\begin{equation} \label{LandZen}
r=e^{-\frac{\pi v^2}{2\alpha}}
\end{equation}
expressing the state population changes in terms of the rate $\alpha$ at which
the diagonal terms of the linear Hamiltonian change their value, and of the
off-diagonal interaction strength $v$~\cite{landzen}. It must be noted that
since we are considering two states only, $v$ represents the energy gap, too.
The transition probability is symmetric in the linear case,
i.e. the tunneling rate from the lower level to the upper one is the same as in
the opposite direction.

\begin{figure}[htbp]
\centering\begin{center}\mbox{\epsfxsize 3.2 in
\epsfbox{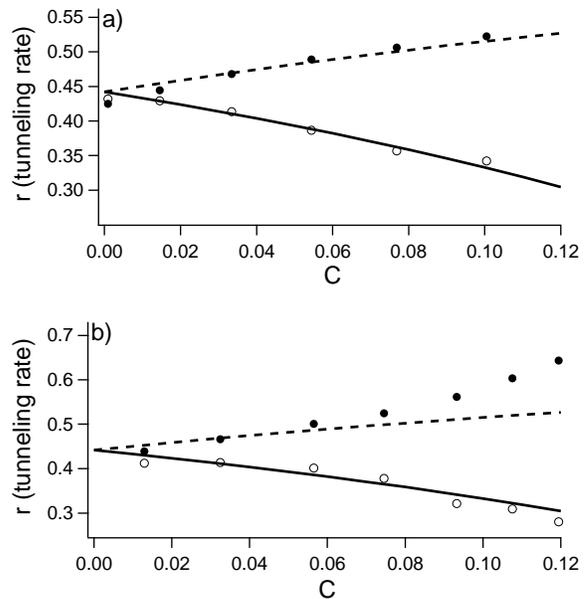}}
\caption{LZ tunneling rate $r$ as a function
of the nonlinear parameter $C$. The dashed and continuous lines represent the
results of the two-level model  of eq.~\eqref{two-state-eq} for a transition
from the lower level to the upper one and vice versa, respectively. The open
and filled symbols are the results of the numerical integration of
eq.~\eqref{schrod} taking into account the presence of the harmonic trap and
simulating the full experimental procedure described in
section~\ref{experiment}. Transition rates were evaluated by means of the
Fourier transform method. Results are for $v=0.1375$ corresponding to $s=2.2$
and $\alpha=0.03636$ corresponding to $a_{L}=2.925 \mathrm{m\,s^{-2}}$. $C$ was
varied in (a) by varying the trap frequency, and  in (b) by varying the atomic
scattering length
$a_{s}$,
the harmonic trap frequency being fixed at 20 Hz.}\label{RvsC}
\end{center}\end{figure}

In the nonlinear regime, as the nonlinearity parameter $C$ grows, the lower to
upper tunneling probability grows as well until an adiabaticity breakdown
occurs at $C=v$ where the swallow tail structure appears~\cite{wuniu-lz}. The 
upper to lower tunneling probability, on the other hand,
decreases with increasing nonlinearity~\cite{note2}. We derive the tunneling
rate from the numerical integration of eq.~\eqref{two-state-eq}. In
fig.~\ref{RvsC} we plot the lower to upper tunneling rate (initial
$(a,b)=(1,0)$ in eq.~\eqref{twolevels}) and the upper to lower tunneling rate
(initial $(a,b)=(0,1)$ in eq.~\eqref{twolevels}) of the Bose-Einstein
condensate as a function of the nonlinear parameter $C$. We see that for $C=0$
the rate is the same for both tunneling directions whereas for $C\neq 0$ the
two rates are different. We confirm the presence of tunneling asymmetry by
integrating eq.~\eqref{schrod} directly (taking into account the full
experimental protocol), finding qualitative agreement with the
prediction of the two-state model.

From an analytical point of view
the nonlinear regime is interpreted straightforwardly by writing
eq.~\eqref{two-state-eq} as
\begin{multline} \label{two-state1-eq}
i\frac{\partial}{\partial t}\,\begin{pmatrix} a \\ b \end{pmatrix}
=
\left[\alpha t\frac{\sigma_{3}}{2}+v\frac{\sigma_{1}}{2}\right]
\begin{pmatrix} a \\ b \end{pmatrix} + \\
+ \frac{C}{2} \begin{pmatrix} 0 & 2b^*a \\ 2a^*b & 0
\end{pmatrix} \begin{pmatrix} a \\ b \end{pmatrix}
\end{multline}
This different form leads to the same condensate dynamics since 
the difference between the matrices in eq.~\eqref{two-state1-eq} and
eq.~\eqref{two-state-eq} times the $(a,b)$ vector is equivalent to the identity
operator. The difference is then an offset in the energy of the system. In
eq.~\eqref{two-state1-eq} we identify an off-diagonal term $|v+2C\,a^*b|$
acting as an effective potential $v_{eff}$; the modulus is needed since $a^*b$
is complex. The off-diagonal scalar product between the two states $a^*b$ may
be evaluated using the adiabatic approximation technique~\cite{crisp}
\begin{equation}
a^*b=-\frac{v}{2\sqrt{\alpha^2 t^2
+v^2}}e^{i\frac{\alpha}{\alpha^2 t^2 +v^2}}
\label{outdiagonal}
\end{equation}
for a transition from the lower state to the upper state. For tunneling in the
opposite direction, $a^*b$ simply changes sign.  
The explicit expression for $v_{eff}$ is
\begin{equation}
v_{eff}=v\sqrt{1\pm\frac{2C}{\sqrt{\alpha^2
t^2 +v^2}}\cos\left(\frac{\alpha}{\alpha^2 t^2 
+v^2}\right)+\frac{C^2}{\alpha^2 t^2 +v^2}}.
\end{equation}
Within the spirit of the adiabatic approximation we put $\alpha=0$ obtaining
\begin{equation} \label{effectmattia}
v_{eff}=v\sqrt{1\pm\frac{2C}{v}+\frac{C^2}{v^2}}=v\left(1\pm\frac{C}{v}\right)
\end{equation}
where the upper and lower signs corresponds to initial conditions of
excited/ground states.
For the ground energy band the following effective potential was  introduced by
Choi and Niu~\cite{choi99} and experimentally tested by Morsch {\it et
al.}~\cite{morsch01}:
\begin{equation}
     v_{eff}=\frac{v}{1+4C}
\label{effectchoi}
\end{equation}

For small $C$ values we modify
the LZ formula of eq.~\eqref{LandZen} to include nonlinear corrections,
replacing the potential $v$ by the effective potential $v_{eff}$~\cite{note4}.

\begin{figure}[htbp]
\centering\begin{center}\mbox{\epsfxsize 3.5 in
\epsfbox{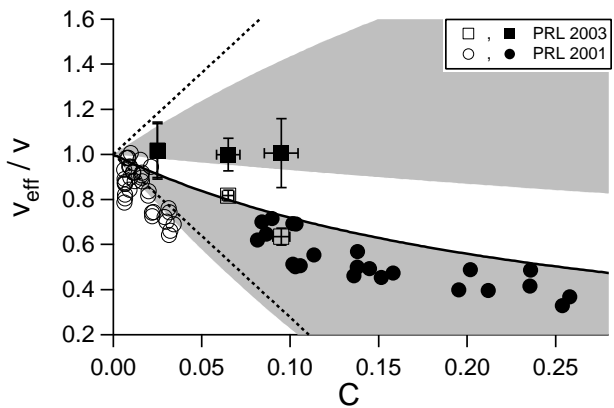}}
\caption{
Ratio between the effective potential $v_{eff}$ and the bare potential $v$ as a
function of the nonlinear parameter $C$. The dotted lines are
eq.~\eqref{effectmattia} for both tunneling directions. The solid line is
eq.~\eqref{effectchoi} for tunneling from ground to excited band. The shaded 
regions are presented in the text. The squared marks 
are the experimental values
calculated using the LZ tunneling rate of eq.~\eqref{LandZen} and data
from~\cite{jona1} for a transition from the lower band to the upper band (open
marks) and in the opposite direction (filled marks).  The circular marks are
data from~\cite{morsch01} with optical lattice step $d=1.18\,\mathrm{\mu m}$
(filled marks) and $d=0.39\,\mathrm{\mu m}$ (open marks), as explained in
section~\ref{experiment}. All calculations and experimental data are for
$v=0.1375$ corresponding to $s=2.2$ and $\alpha=0.03636$ corresponding to
$a_{L}=2.925 \mathrm{m\,s^{-2}}$.}
\label{veff_v0_data}
\end{center}\end{figure}

Both expressions~\eqref{effectmattia} and~\eqref{effectchoi} assume to use  an
effective potential taking into account the nonlinearity of the system. This
assumption  has some validity limits for the potential $v$ and/or the
nonlinearity strength $C$. Equation~\eqref{effectchoi}, derived using
perturbation theory, is expected to hold as long as the condensate density is
nearly uniform (i.e. $v_{eff} \ll 1$). This condition may be realized with
either a weak external potential or a strong atomic interaction~\cite{choi99}.
Because we consider a condensate weakly perturbed by the presence of the
periodic potential, we expect eq.~\eqref{effectchoi} to hold in a neighborhood
of the center of the band where the effects of the periodic potential are
weakest and the band resembles the free particle dispersion curve. On the other
hand we developed the two level model by approximating the wave function at the
band edge; hence we expect eq.~\eqref{effectmattia} to hold in a neighborhood
of the band edge.

In order to test our intuitions we performed a numerical variational solution
of eq.~\eqref{schrod-adim} based on the Mathieu functions, which  are the exact
solutions of the linear problem (i.e. eq.~\eqref{schrod-adim} with $C=0$ and 
substituting the time derivative with the chemical potential). We evaluated the
chemical potential using the Mathieu functions and we minimized this value by
varying the parameter in the Mathieu function corresponding to the lattice
potential strength, thus obtaining an effective potential. In the effective
potential description, this is the best approximation of the {\em exact}
solution of the nonlinear problem. We repeated this calculation for different
values of the atomic quasimomentum $q$, from $q=0$ to the band edge $q=1/2$.
Our findings are summarized in fig.~\ref{veff_v0_data} where the lower and
upper shaded areas correspond to intermediate values of the quasimomentum in
both the ground band (lower shaded area) and excited band (upper shaded area).
The effective potential at $q=0$ corresponds to the upper edge of the lower
shaded area (ground band) and to the lower edge of the upper shaded area
(excited band). Figure~\ref{veff_v0_data} also reports the results of
eq.~\eqref{effectmattia} and eq.~\eqref{effectchoi}.
Equation~\eqref{effectchoi} is indistinguishable from the upper edge of the
lower shaded area, hence we infer that it can only hold for the lower band
exactly at quasimomentum $q=0$, in agreement with our previous analysis. On the
other hand, because eq.~\eqref{effectmattia} reproduces very well the lower
edge of the lower shaded area and the upper edge of the upper shaded area, we
confirm its validity range to be limited to a neighborhood of the band edge.

Equations~\eqref{effectmattia} and~\eqref{effectchoi} define a different role
of the nonlinearity. Equation~\eqref{effectchoi} states that the ratio
$v_{eff}/v$ does not depend on the potential magnitude $v$, but only on the
magnitude of the nonlinear parameter $C$, whereas eq.~\eqref{effectmattia}
predicts a deviation from the linear case according to the ratio between the
magnitude of the nonlinearity and the bare potential strength. In  the present
description,  the energy scale determined by the bare potential strength $v$
appears to be irrelevant at the center of the band; the only relevant energy
scale is fixed by the mean field interaction strength $C$. On the contrary, the
energy scale determined by $v$ acquires more and more significance toward the
band edge. At the band edge the two energy scales have the same relevance and
$v_{eff}/v$ only depends on the ratio of them. The swallow tail threshold
$C/v=1$ represents a critical value for the system since the effective
potential~\eqref{effectmattia} vanishes for a transition from the lower band to
the upper band while it doubles for a transitions in the opposite direction.
Therefore when $C/v > 1$, it is no longer possible to interpret the
nonlinearity  through an effective potential.

The difference in the tunneling rates in the two directions may be  also
derived from the difference between the energy gaps $\Delta \mu_{10}$ and
$\Delta \mu_{01}$ (fig.~\ref{gap}a) using eq.~\eqref{LandZen} to link  the
transition rates to the energy gaps (fig.~\ref{gap}(b)).  As a consequence, the
tunneling rate is enhanced in one case and it is suppressed in the other one. 
These tunneling  rates are in very good  agreement with those predicted by the
effective potential of eq.~\eqref{effectmattia}.

\subsection{Experimental observations}
Landau-Zener tunneling between the two lowest energy bands of a condensate
inside an optical lattice was investigated in the following way (see
fig.~\ref{Band-structure}). Initially, the condensate was loaded adiabatically
into one of the two bands, either in the ground band or into the excited band.
Subsequently, the lattice was accelerated in such  a way that, at  the BZ edge,
a finite probability for tunneling into the other band resulted. After the 
tunneling event, the two bands had populations reflecting the Landau-Zener
tunneling rate. In order to experimentally determine the number of atoms in the
two bands, we then {\em increased} the lattice depth  and {\em decreased} the
acceleration. In this way, successive crossings of the band edge resulted in a
much reduced Landau-Zener tunneling probability between the ground band and the
first excited band (of order a few percent), as illustrated in
fig.~\ref{Band-structure}. The fraction of the condensate that populated the
ground band after the first tunneling event, therefore, remained in that band,
whereas the population of the first excited band  underwent tunneling to the
second excited band with a large probability (around $90$ percent) as the gap
between these two bands is smaller than the gap between the two lowest bands.
Once the atoms underwent tunneling into the second excited band, they
essentially behaved as free particles.  For both tunneling directions, the
tunneling rate $r$ was derived from the  ratio between the number of atoms
experiencing the tunneling process  and  the total number of atoms measured
from the absorption picture.

In order to verify the experimental procedure for measuring the tuneling rate
in both directions and also to verify that in the linear regime the
Landau-Zener tunneling was symmetric, we measured  the two tunneling rates as a
function of the lattice depth for a condensate in a weak magnetic trap and
hence a small value of the interaction parameter $C$. In this case, both
tunneling rates were essentially the same and agreed well with the linear
Landau-Zener prediction, as reported in~\cite{jona1}. By contrast, when $C$ was
increased, the two tunneling rates began to differ. For instance for the
parameter $C=0.095(9)$, the measured tunneling rate from the ground to excited
state was $r=0.72\pm 0.10$, whereas we measured $r=0.37\pm 0.05$ in the
opposite direction, proving the tunneling asymmetry. 

The effective potential description allows us  to present within a unified
picture the nonlinear Landau-Zener tunneling rates measured in
refs.~\cite{morsch01,jona1},  as  plotted in fig.~\ref{veff_v0_data}, together
with the theoretical  predictions. We derive a qualitative agreement with the
theoretical predictions of the non-linear Landau-Zener model, whereas
quantitatively there are significant deviations. We believe these to be partly
due to experimental imperfections. In particular, the sloshing (dipolar
oscillations) of the condensate inside the magnetic trap can lead to the
condensate not being prepared purely in one band due to non-adiabatic mixing of
the bands if the initial quasimomentum is too close to a band-gap.  Furthermore
the amplitude of the shaded areas in fig.~\ref{veff_v0_data} points out a
strong dependence of the effective potential on the quasimomentum. If the
condensate trapped in the periodic potential has a finite extension in momentum
space, the overall effect of the effective potential on the tunneling
probability is not restricted to a single quasimomentum but represents a mean
effect over all the quasimomenta of the condensate.  These effects could be
responsible for the experimental points not falling exactly on the lines
corresponding to the effective potential evaluated at the band edge. In any
case the experimental data must fall within the shaded areas corresponding to
the band in which the condensate was loaded, as in fig.~\ref{veff_v0_data}.

Because of the elastic force of the magnetic harmonic trap, it is important in
the experiment not to drag the condensate too far from the rest position. If
the condensate is dragged too far, the dragged part starts to feel the
restoring force due to the harmonic potential and hence does not feel a {\em
constant} force anymore. For this reason it is not possible to study large $C$
values by varying only the harmonic trap frequency.  For our experimental
parameters, $C\approx 0.11$ was found to be the largest acceptable value,
corresponding to a harmonic trap frequency of about $50\,\mathrm{Hz}$. 
Furthermore, a numerical simulation of the experiment showed that for large
values of $C$, for which the magnetic trap frequency was large, the measured
tunneling rates were significantly modified by the presence of the trap.
However, we verified in the simulation that when $C$ was varied without varying
the trap frequency, the asymmetric tunneling effect persisted. In future
experiments, one might study large $C$ values by increasing the atomic density
in the condensate by using an additional optical trap, in order to increase the
radial trapping frequency or, alternatively, by using Feshbach resonances to
vary the atomic scattering legth $a_s$. The results of a numerical simulation
using the latter method are reported in fig.~\ref{RvsC} (b).

\section{Instabilities}
\label{instab}
\subsection{Experimental observations}
\begin{figure}[htbp]
\centering\begin{center}\mbox{\epsfxsize 3.0 in
\epsfbox{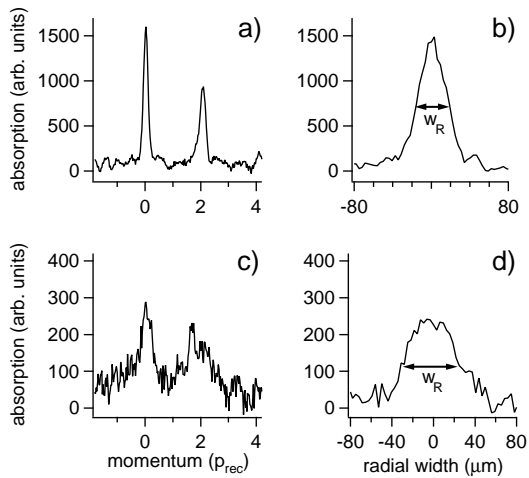}} \caption{Integrated longitudinal
and transverse profiles of the interference pattern of a condensate released
from an optical lattice after acceleration to a quasimomentum $q\approx 0.9
q_{B}$ and a subsequent time-of-flight of $21\,\mathrm{ms}$. The Gaussian width
$w_{R}$ of the transverse width is marked  in (b) and (d). In (a) and (b), the
acceleration $a_{L}$ was $5\,\mathrm{m\,s^{-2}}$, whereas in (c) and (d)
$a_{L}=0.3\,\mathrm{m\,s^{-2}}$. In (a) and (c), the horizontal axis has been
rescaled in units of recoil momenta. Note the different vertical axis scales
(by a factor $4$) for the upper and lower graphs. The total number of atoms was
measured to be the same in
both cases.}\label{peaks}
\end{center}\end{figure}

In the previous experiment we investigated the dependence of the tunneling
probability as a function of the nonlinearity magnitude with fixed
acceleration. We have also investigated the stability of the condensate as a
function of the acceleration with (roughly) fixed
nonlinearity~\cite{cristiani04}.  When the condensate acquired a quasimomentum 
close to the band edge, the  unstable solutions of eq.~\eqref{schrod} grow
exponentially in time, leading to a loss of  phase coherence of the condensate
along the direction of the optical lattice. In our experiment, the time the
condensate spent in the `critical region' where unstable solutions existed, was
varied through the lattice acceleration. When the acceleration was small the
condensate moved across the Brillouin zone more slowly and hence the growth of
the unstable modes~\cite{wu03} became more important. Figs.~\ref{peaks}c
and~\ref{peaks}d show typical integrated profiles of the interference pattern
obtained for a lattice acceleration $a_{L}=0.3\,\mathrm{m\,s^{-2}}$. Here, the
condensate reached the same point close to the Brillouin zone edge as in
Figs.~\ref{peaks}a and~\ref{peaks}b, but because of the longer time it
spent in the unstable region, the interference pattern was almost completely
washed out. It is also evident that the radial expansion of the condensate was 
considerably enhanced when the Brillouin zone was scanned with a small
acceleration.

\begin{figure}[htbp]
\centering\begin{center}\mbox{\epsfxsize 3.0 in
\epsfbox{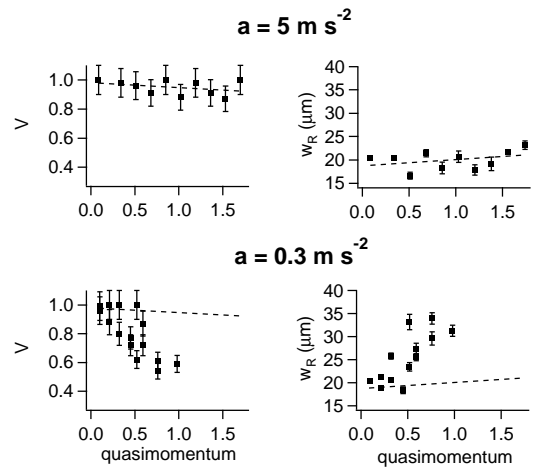}}
\caption{Visibility $V$ and radial width $w_{R}$ as a function of the
final quasimomentum (in units of $q_{B}$) for different accelerations. As the
acceleration was lowered, instabilities close to the BZ edge decreased the
visibility and increased  the radial width. For comparison, in each graph the
(linear) fits to $V$ and $w_{R}$ for the $a_{L}=5\,\mathrm{m\,s^{-2}}$ data are
included. }\label{instability}
\end{center}\end{figure}

In order to characterize more quantitatively our experimental findings on the
instability, we defined two observables for the time-of-flight interference
pattern. By integrating the profile in a direction {\em perpendicular to} the
optical lattice direction, we obtained a two-peaked curve (see fig.~\ref{peaks}
(a)) for which we defined a visibility $V$ (in analogy to spectroscopy)
reflecting the phase coherence of the condensate. $V$ is close to $1$ for
perfect coherence, whereas $V\longrightarrow 0$ for an incoherent condensate. 
The second observable we defined, was the width $w_{R}$ of a Gaussian fit to
the interference pattern  integrated {\em along} the lattice direction over the
extent of one of the peaks (see fig.~\ref{peaks} (b) and (d)).

In~\cite{cristiani04} we measured $V$ and $w_{R}$ as a function of the final
quasimomentum value reached for different values of the acceleration.
Figure~\ref{instability} shows clearly that for large accelerations, both $V$
and $w_{R}$ remained reasonably stable when the edge of the Brillouin zone is
crossed. In contrast, for $a_{L}=0.3\,\mathrm{m\,s^{-2}}$ one sees a drastic
change in both quantities as the quasimomentum approaches the value $q_{B}$.
For those accelerations, the condensate spent a sufficiently long time in the
BZ unstable region  and hence lost its phase coherence, resulting in a sharp
drop of the visibility. At the same time, the radial width of the interference
pattern increased. This increase is evidence for an instability in the
transverse directions. For $a_{L}=0.3\,\mathrm{m\,s^{-2}}$, the interference
patterns for quasimomenta larger than unity were so diffuse that it was not
possible to measure either the visibility or the radial width in a meaningful
way.

In~\cite{cristiani04} we estimated an instability growth rate of
$10^3\,\mathrm{s^{-1}}$ from the time spent by the condensate in the unstable
region of the Brillouin zone. This experimental value agrees with the
theoretical prediction of $2000\,\mathrm{s^{-1}}$ estimated by Niu~\cite{wu03}.
Fallani {\it et al.}~\cite{fallani04} carried out an experiment for operating
conditions similar to those of our investigations, but measured the effect of
the instability through the atom loss rate from the condensate. They reported
atom loss rates in the range $40-60\,\mathrm{s^{-1}}$. Although it is not clear
how to link the atom loss rate to the instability growth rate, the qualitative
agreement (i.e. the dependence of both quantities on the quasimomentum) between
their experimental results and their theory is good. Their experiment was
subsequently simulated by Modugno {\it et al.}~\cite{modugno04} who predicted
growth rates of the unstable modes in the range $2000-3000\,\mathrm{s^{-1}}$.
In that work the growth rate is expressed in units of the transverse magnetic
trap frequency. This choice could be interpreted as meaning that the
instability depends on the radial modes. It has, however, been clarified by one
of the authors of that paper~\cite{dalfovopc} that this is not the case and
that the choice of the transverse frequency as the frequency unit was made
purely for reasons of convenience. In fact, the dependence of the growth rate
on the transverse frequency needs to be studied in more depth. We should point
out that all the theoretical analyses predict the occurrence of the
instabilities at the microscopic level only, while the experimental
observations reflect the macroscopic changes of the condensate features, for
instance the visibility in our case or the atom loss rate in the case
of~\cite{fallani04}.

\begin{figure}[htbp]
\centering\begin{center}\mbox{\epsfxsize 3.0 in
\epsfbox{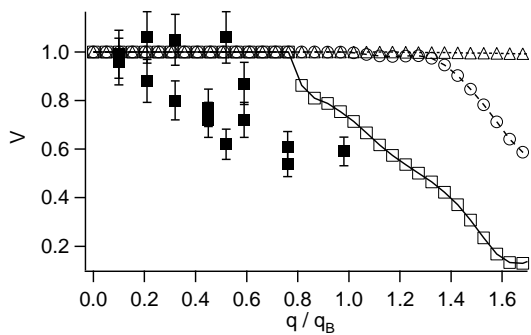}}
\caption{Results of a one-dimensional numerical simulation for the
visibility $V$ versus the condensate final quasimomentum $q$ in the conditions
of  the  experiment for  acceleration $a=0.3\,\mathrm{m\,s^{-2}}$ and for
different values of the nonlinear parameter $C$. The open  squares, circles and
triangles correspond to $C=0.008$ (the value for our experiment), $C=0.004$ and
$C=0$, respectively.  The dashed lines connect the theoretical points to guide
the eye.  The closed symbols are the experimental values of the visibility as
reported in fig.~\ref{instability} for
$a=0.3\,\mathrm{m\,s^{-2}}$.}\label{instab_theo}
\end{center}\end{figure}

\subsection{Modelling} 
We compared our experimental results to a simple 1-D numerical simulation.
Figure~\ref{instab_theo} shows the results of a numerical integration of the
one-dimensional Gross-Pitaevskii equation with the parameters of our
experiment. The visibility was calculated in the same way as was done for the
experimental interference patterns. It is clear from this simulation that it
is, indeed, the nonlinearity that is responsible for the instability at the
edge of the Brillouin zone. When $C$ was set to $0$ in the numerical
simulation, the visibility remains unaltered when the BZ edge is crossed,
whereas for finite values of $C$ the visibility decreases as the quasimomentum
$q_{B}$ was approached. Furthermore, the larger the value of $C$, the more
pronounced was the decrease in visibility near the band edge. For $C=0.008$,
corresponding to the value realized in our experiment, the onset of the
instability was located just below a quasimomentum of $0.8q_{B}$.
Experimentally, we found that the visibility started decreasing consistently
beyond a quasimomentum of $\approx 0.6-0.7\,q_{B}$, agreeing reasonably well
with the results of the simulation. The presence of experimental points with
visibility less than unity in the quasimomentum region below $0.6q_{B}$ can be
explained by considering the initial sloshing of the condensate inside the
harmonic trap. This oscillation introduced a sensitive error in the
quasimomentum determination which was estimated to be of the order of
$0.3q_{B}$.

The determination of the quasimomentum corresponding to the onset of
instability is a subject that has been examined in the literature, and in a
recent experimental investigation the quasimomentum scanning across the
Brillouin zone was stopped at different values in order to verify the
instability growth~\cite{fallani04}. Our measured values for the instability
onset are in agreement with those measured in that reference.

\section{Conclusions}\label{conclusions}
We have numerically simulated and experimentally studied the dynamics of a Bose
Einstein condensate inside a periodic potential in two different regimes. In
the  Landau-Zener regime, we investigated the tunneling between two energy
bands in a periodic potential and found that, in the presence of a nonlinear
interaction term, an asymmetry in the tunneling rates arises. Experimentally,
we measured these tunneling rates for different values of the interaction
parameter and found qualitative agreement with the simulations.  In the 
instability regime, we studied the stability of a BEC in the vicinity of the
band edge, finding good agreement between experimental results and the
theoretical expectation of unstable behavior. These observations confirmed that
Bose-Einstein condensates may be used to simulate  a variety of nonlinear
physics configurations. Future experiments could probe the complicated and
time-dependent tunneling behaviour due to the changing tunneling rate for
multiple crossings of the zone edge.

To conclude, we note that the phenomenon of asymmetric tunneling should be a
rather general feature of quantum systems exhibiting a nonlinearity. For
instance, calculating the energy shift due to a nonlinearity for two adjacent
levels of a harmonic oscillator, one finds that both levels are shifted upwards
in energy, the shift being proportional to the population of the respective
level. The energy difference between the levels, therefore, decreases if only
the lower state is populated and increases if all the population is in the
upper level. Furthermore the asymmetric Landau-Zener tunneling rate was applied
to intepret the photoassociation of a Bose-Einstein
condensate~\cite{ishkhanyan04}, with the surprising  result that at small
crossing rates the no-transition probability is directly proportional to the
rate at which the resonance is crossed.

We acknowledge financial contributions from the ``Progetto MIUR COFIN-2004''.


\begin{thebibliography}{99}
\bibitem{note1} Obviously there is no need for the two states to be  the ground
and first excited states. It is only required for the two states to be
adjacent.
\bibitem{note2} eq.~\eqref{two-state-eq} predicts that for or  occupation with
equal amplitude of the lower and upper states $|a|^{2}=|b|^{2}$, the nonlinear
corrections leading to the asymmetric tunneling play no role.
\end{thebibliography}
\end{document}